\pdfoutput=1

\documentclass[conference,a4paper]{IEEEtran}

\usepackage[utf8]{inputenc} 
\usepackage[T1]{fontenc}
\usepackage{url}              
\usepackage{cite}             

\usepackage[cmex10]{amsmath}  
\usepackage{mleftright}       
\mleftright                   

\usepackage{graphicx}         
\usepackage{booktabs}         
 \usepackage[caption=false,font=footnotesize]{subfig}

\newtheorem{theorem}{Theorem}
\newtheorem{definition}{Definition}
\newtheorem{corollary}{Corollary}
\newtheorem{lemma}{Lemma}
\newtheorem{remark}{Remark}
\newtheorem{proposition}{Proposition}

\usepackage{tikz}
\usetikzlibrary{shapes,snakes}
\begin{document}

\title{Joint Design of Sampler and Compressor for Timely Status Updates: Age-Distortion Tradeoff} 
\author{Jun Li,
  	Wenyi Zhang
}

\maketitle

\begin{abstract}
  We consider a joint sampling and compression system for timely status updates. 
  Samples are taken, quantized and encoded into binary sequences, which are sent to the destination. 
  We formulate an optimization problem to jointly design sampler, quantizer and encoder, minimizing the age of information (AoI) on the basis of satisfying a mean-squared error (MSE) distortion constraint of the samples.
  We prove that the zero-wait sampling, the uniform quantization, and the real-valued AoI-optimal coding policies together provide an asymptotically optimal solution to this problem, i.e., as the average distortion approaches zero, the combination achieves the minimum AoI asymptotically. 
  Furthermore, we prove that the AoI of this solution is asymptotically linear with respect to the log MSE distortion with a slope of $-\frac{3}{4}$.  
  We also show that the real-valued Shannon coding policy suffices to achieve the optimal performance asymptotically.
  Numerical simulations corroborate the analysis.
\end{abstract}

\section{Introduction}
\label{sec:double-blind-policy}


In real-time data compression and transmission systems, it is desirable that the destination can reconstruct the information generated by the source as early as possible. 
In fact, due to the rapid increase in the amount of data and the requirement for application timeliness, if the entire data stream is encoded and transmitted with a high degree of accuracy, the information will become stale. 
Therefore, for such sources, we first need to select the appropriate information, and then design a high-performance compressor for the timeliness of information.
However, if we reduce the information staleness significantly, the distortion between the original symbols and the received will increase greatly. This shows that there is an important tradeoff between freshness and distortion of the information. 
Hence, on the basis of satisfying the distortion constraint, how to make the information fresher is a critical task for the real-time application. 
This type of application will be common, for example, in autonomous vehicle applications, where we need to transmit the relatively accurate real-time location information to the control center timely so that the controller can make corresponding decisions as quickly as possible.

It is well known that age of information (AoI) \cite{Kaul_2012} is considered to capture the core notion of timeliness, and it has been widely used in communication systems. 
Related works include \cite{Sun_2017,Sun_2019,Zhong_2016,Abend_2021,Zhong_2017,Mayekar_2020,Bastopcu_2020,Sun_2020,inan2021optimal}. 
The optimal sampling policy for generate-at-will sources have been investigated in \cite{Sun_2017} and \cite{Sun_2019} without involving coding.  Real-time fixed-to-variable, variable-to-fixed and variable-to-variable lossless source coding schemes have been studied in \cite{Zhong_2016}, \cite{Abend_2021} and \cite{Zhong_2017} respectively. In \cite{Abend_2021,Zhong_2016,Zhong_2017}, all the symbols generated are transmitted. 
Unlike \cite{Abend_2021,Zhong_2016,Zhong_2017}, in \cite{Mayekar_2020}, \cite{Bastopcu_2020} symbols are encoded and transmitted once the previous transmission is completed, which is called the zero-wait policy. 
In \cite{Mayekar_2020}, a variational formula has been proposed to obtain the optimal AoI code. In \cite{Bastopcu_2020}, the partial update where the transmitter combines the realization of updates to reduce the AoI has been studied. Nevertheless, these works do not consider the analog (i.e., continuous-valued) source sequences. Moreover, although \cite{Mayekar_2020,Bastopcu_2020} studied source coding under the zero-wait policy, their schemes are not optimal unless the service time satisfies certain technical conditions in \cite{Sun_2019}. Furthermore, the tradeoff between freshness and distortion is considered in \cite{Sun_2020} from the remote estimation perspective. In \cite{inan2021optimal}, an
age-distortion tradeoff has been discussed, where the distortion quantifies the importance of data. 

\begin{figure*}[htbp]
	\centering
	\begin{tikzpicture} 
			\node[draw,circle,fill=none,text=black] (Source) at (0.5,0) {Source};
			\node[draw,minimum height=1cm,minimum width=2cm] (Sampler) at(3.5,0) {Sampler};
			\node[draw,minimum height=1cm,minimum width=2cm] (Quantizer) at(6.5,0) {Quantizer};
			\node[draw,align=center,minimum height=1cm,minimum width=2cm] (Encoder) at(9.5,0) {Encoder};
			\node[draw,align=center,minimum height=1cm,minimum width=2cm] (Noiseless channel) at(13.5,0) {Noiseless\\ channel};
			\node[draw,align=center,minimum height=1cm,minimum width=2cm] (Receiver) at(16.5,0) {Receiver};
			\draw [dashed](2,1)  rectangle (11,-1) ;
			\node(A) at(6.5,-1.25) {Transmitter};
			\node(B) at(6.5,0.87){} ;
			\draw[->] (Source)-- (Sampler)node[near start,above] {$\quad X_t$}; 
			\draw[->] (Sampler)--(Quantizer)node[near start,above] {$\quad \ \ X_{t_k}$};
			\draw[->] (Quantizer)--(Encoder)node[near start,above] {$\quad \ Q(X)$};
			\draw[->] (Encoder)--(Noiseless channel)node[near start,above] {$\quad \quad \quad \quad \ 01101...$};
			\draw[->] (Noiseless channel)--(Receiver);
			\draw[->] (Receiver) |- +(0,1.5) -|(B) node[near start,above] {ACK};
	\end{tikzpicture}
	\caption{System model}
	\label{fig:Fig.1}
\end{figure*}
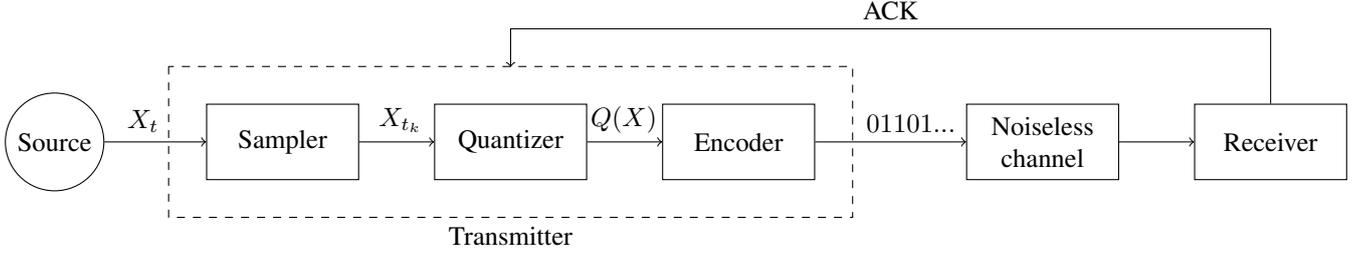

In this paper, we consider a joint sampling and compression system for a generate-at-will analog source, as shown in 
Fig. \ref{fig:Fig.1}. The sampler decides the generation time of the updates and the obtained updates are quantized into discrete random symbols which are assigned variable-length prefix-free codewords by the encoder. 
Then the transmitter sends these codewords through a noise-free channel that can transmit one bit per unit time. 
We formulate an optimization problem to jointly design the sampler, quantizer and encoder to minimize the AoI given a distortion constraint of the samples.

Our main contributions are as follows:
We find an asymptotically optimal joint policy, consisting of the zero-wait sampling, the uniform quantization and the real-valued AoI-optimal coding policies, to achieve the minimum of the optimization problem asymptotically.
We find tight bounds of the optimal AoI to prove the results. This method can avoid solving an optimization problem directly.
At the same time, by using this method, we can also prove that the performance curve of the optimal AoI versus log distortion is a linear function with a slope of $-\frac{3}{4}$ asymptotically as the distortion approaches zero
and that the real-valued Shannon code is sufficient to achieve the optimal performance asymptotically.

The remaining part of this paper is organized as follows. In Section \uppercase\expandafter{\romannumeral2}, we describe the system model and formulate the joint sampling and compression problem.  In Section \uppercase\expandafter{\romannumeral3}, we present the asymptotically optimal joint policy that minimizes the AoI and analyze its performance. Some numerical results are provided in Section \uppercase\expandafter{\romannumeral4}. Finally, we conclude the paper in Section \uppercase\expandafter{\romannumeral5}.

%
%

\section{System model}
\label{sec:page-limit}

We consider a continuous-time status update system with a single generate-at-will source (i.e., the update symbols can be generated at any time), as depicted in \figurename \ \ref{fig:Fig.1}. In practice, we can use the continuous-time system to model  status update systems with a high update rate \cite{Sun_2019}.
At each time instant, the source generates independent and identically distributed (i.i.d.) symbols $X$ with a known probability density function (pdf) $f(x)$. 
We assume that the pdf $f(x)$ satisfies the following two conditions:

Condition A: $f(x)$ is continuous, differentiable, and has a bounded support interval $I$. The maximum of $f(x)$ is denoted by $M$.

Condition B: The integral
$\int_I f(x)\log_2^2 {(f(x))}dx$ and the differential entropy $h(X)=\int_I f(x)\log_2 {(f(x))}dx$ exist. 

A sampler takes samples from the source. The time-stamped samples are fed into a variable-rate
quantizer and are then assigned binary prefix-free codewords by an encoder.
We also assume that the transmitter can obtain the instantaneous channel idle/busy states. 
The symbols encoded are sent through a noise-free channel to the receiver one bit per unit time, only when the channel is free. That is, the service time of transmitting a symbol is equal to the codeword length assigned to this symbol. 
The system consists of two main parts: an age-based sampler and an age-based compressor which further consists of a quantizer and an encoder.

We use the metric of AoI to measure the freshness of the information at the destination. Let $U_t$ be the generation time of the most recently delivered symbol by time $t$. Let $\Delta_t$ be the instantaneous age at time $t$, which is defined by \cite{Kaul_2012}
\begin{align}
	\Delta _t=t-U_t.
\end{align}
The AoI, i.e., the average age, is defined by
\begin{align}
	{\rm AoI}=\limsup_{T\to \infty} \frac{1}{T}E\bigg[\int_{0}^{T} \Delta_t dt \bigg].
\end{align}

We describe the two components of this system in detail below.

%
%
%
%
\subsection*{A. Age-based Sampling}
We refer to a series of generation times of updates as a sampling policy denoted by $S$. Each sampling time is chosen based on history and current information of the idle/busy state of the channel. 
If a compressor is given, we will know the quantization distortion of the source symbols as well as the distribution of the service time. We need to find a sampling policy to minimize the AoI, a problem similar to those treated in \cite{Sun_2017} and \cite{Sun_2019}. The problem is formulated as follows:
\begin{align}
	{ {\rm AoI}^*}=& \min_{S\in \mathcal{S}} \limsup_{T\to \infty} \frac{1}{T}E\bigg[\int_{0}^{T} \Delta_t dt \bigg],
\end{align}
where $\mathcal{S}$ denotes the set of all sampling policies.
If the service times are i.i.d., it has been shown that
\begin{align}\label{AoI}
	{\rm AoI}=\frac{E[(L+Z(S))^2]}{2E[L+Z(S)]}+E[L],
\end{align}
and we only need to consider stationary deterministic sampling policies \cite{Sun_2017}. Let $\mathcal{S}_\mathrm{d}$ denote the set of stationary deterministic policies. A waiting time $Z\in [0,W]$ is inserted before the generation of a new symbol, which is a nonnegative deterministic function of the service time, i.e. the codeword length assigned to the symbol. In this case, a sampling policy is equivalent to determining a series of waiting times denoted by $S=(Z_1,Z_2,...)$. A special sampling policy is the zero-wait policy whose waiting time is always zero, i.e., $S_{\mathrm{z}}=(0,0,...)$.
\subsection*{B. Age-based Compression}
Given a sampling policy $S$, the AoI is determined by the service time of the symbols (i.e. the lengths of assigned codewords to the quantization cells). For a quantizer $Q$, we use ${a_i}$ to denote the $i$th quantization interval endpoint. Each interval $[a_{i-1},a_i]$ is represented by a representation point $c_i$, whose occurrence probability is denoted by $p_i$. 
Let $\mathcal{Q}$ denote the set of quantization policies. The mean-squared distortion of $Q$ is given by
\begin{equation}
	D(Q)=\sum_i \int_{a_{i-1}}^{a_i} (x-c_i)^2 f(x)dx.  
\end{equation}

We assign binary prefix-free codewords to the quantization cells and the codeword length assigned to the $i$th cell is denoted by $l_i$. 
$L$ is a random variable representing the codeword length for the random quantized symbol and $\mathcal{L}$ denotes the set of codeword length assignments. 
We know that prefix-free codes exist for any set of codeword lengths satisfying the Kraft inequality, i.e., $\sum_i 2^{-l_i}\le 1$ \cite{cover1999elements}.
When a sampling policy $S$ and a quantization policy $Q$ are given, we formulate the AoI minimization problem as
\begin{align}
	\min_{\{L\in\mathcal{L}\}}&\frac{E[(L+Z(S))^2]}{2E[L+Z(S)]}+E[L] \\\notag
	\mbox{s.t.}& \sum_i 2^{-l_i}\le 1 \\\notag
	&l_i\in Z^+.
\end{align}

For simplicity, in the subsequent analysis we ignore the integer constraint and consider real-valued length assignments \cite{Mayekar_2020}. The relaxed optimization problem becomes
\begin{align}\label{coding}
	\min_{\{L\in\mathcal{L}\}}&\frac{E[(L+Z(S))^2]}{2E[L+Z(S)]}+E[L] \\\notag
	\mbox{s.t.}& \sum_i 2^{-l_i}\le 1 \\\notag
	&l_i\in R^+.
\end{align}

Then we present some definitions which will be used throughout.
\begin{definition}
	We refer to the codeword lengths that satisfy the problem (\ref{coding}) as the real-valued AoI-optimal coding policy (or real-valued AoI-optimal code) denoted by $F^*$, and use $F_\mathrm{s}$ to denote the real-valued Shannon coding policy (or real-valued Shannon code), i.e., the assigned codeword lengths satisfying $l_i=-\log_2 p_i$. 
	By using the ceiling function, we can obtain the integer-valued AoI-optimal coding policy (or integer-valued AoI-optimal code) and integer-valued Shannon coding policy (or integer-valued Shannon code) respectively.
	%
	%
\end{definition}

When we consider an analog source,
there exists a tradeoff between ${\rm AoI}$ and quantization distortion. 
That is, the distortion is determined by the quantizer, which also determines the codeword length assignment and thus indirectly affects the ${\rm AoI}$.
Hence, we need to determine both the quantization policy and the corresponding coding policy simultaneously, and this leads to a generalization of 
the problem (\ref{coding}) as follows:
\begin{align} \label{quantization}
	\min_{\{Q\in \mathcal{Q}\}}&\min_{\{L\in\mathcal{L}\}}\frac{E[(L+Z(S))^2]}{2E[L+Z(S)]}+E[L] \\\notag
	\mbox{s.t.} &\sum_i 2^{-l_i}\le 1 \\\notag
	&\sum_i \int_{a_{i-1}}^{a_i} (x-c_i)^2 f(x)dx \le D \\\notag
	&l_i\in R^+.
\end{align}


\subsection*{C. Overall Optimization Problem}
In this paper, we study a joint sampling, quantization and coding problem to optimize the AoI of the status update system subject to an average distortion constraint of the samples, and we aim to find a jointly optimal policy $(S,Q,F)$. The problem is formulated as follows:
\begin{align}\label{joint optimal}  
	&\min_{\{S\in \mathcal{S}_{\mathrm{d}}\}}\min_{\{Q\in \mathcal{Q}\}}\min_{\{L\in\mathcal{L}\}} \frac{E[(L+Z(S))^2]}{2E[L+Z(S)]}+E[L] \\\notag
	&\mbox{s.t.} \sum_i 2^{-l_i}\le 1 \\\notag
	&\quad \ \sum_i \int_{a_{i-1}}^{a_i} (x-c_i)^2 f(x)dx \le D \\\notag
	&\quad \quad l_i\in R^+  \notag.
\end{align}

For the problem (\ref{joint optimal}), we observe that the design of the sampler and the encoder are not independent when the quantization region is given. That is, given an encoder, the distribution of the service time is known. 
We can determine the corresponding sampling policy to make the ${\rm AoI}$ optimal. However, the design of the optimal codeword lengths is also determined by the expression of the ${\rm AoI}$ decided by the sampling policy. In fact, in the general case where only the quantization distortion constraint is given, we need to consider all of the factors above. 
Since the three components are tightly coupled, it is challenging to obtain an exact solution directly. 
\subsection*{D. Preliminaries}
Given a quantizer $Q$, the entropy of the quantizer output is denoted by $H[Q(X)]$. 
A quantizer that attains the minimum entropy for a given quantization distortion $D$ is called the optimal quantizer denoted by $Q^*$.
Let $\delta$ denote the quantization cell size of the uniform quantizer. 
Then we present the definition of asymptotically optimal quantizer.
\begin{definition}
	A quantizer $Q$ that satisfies the following condition 
	\begin{align}\label{uniform}
		\lim_{D \to 0} H[Q(X)]-H[Q^*(X)] = 0
	\end{align} 
	is called an asymptotically optimal quantizer.
\end{definition}

A well-known result in high-resolution quantization theory is that the uniform quantizer denoted by  $Q_{\mathrm{uni}}$ is asymptotically optimal for entropy-constrained quantization \cite{Gish_1968}.  The results are recapitulated as follows: 
\begin{lemma}
	For entropy-constrained quantization, the uniform quantizer is asymptotically optimal, i.e.,
	\begin{align}\label{uniform1}
		\lim_{D \to 0} H[Q_{\mathrm{uni}}(X)]-H[Q^*(X)] = 0.
	\end{align}
	Furthermore, we have
	\begin{equation}
		\lim_{D \to 0}H[Q_{{\mathrm{uni}}}(X)]+\log_2\sqrt{12D}=h(X)
	\end{equation}
	and
	\begin{align}\label{uniform2}
		\lim_{\delta \to 0}H[Q_{{\mathrm{uni}}}(X)]+\log_2 \delta=h(X).
	\end{align}
	We also have
	\begin{align} \label{uniform3}
		\lim_{\delta\to0}\frac{D}{\delta^2}=\frac{1}{12}.
	\end{align}
	Combining (\ref{uniform}) and (\ref{uniform3}), we have
	\begin{align}
		\lim_{\delta \to 0} H[Q_{\mathrm{uni}}(X)]-H[Q^*(X)] = 0.
	\end{align}
	
\end{lemma}




\section{Main results}
\label{sec:paper-format}
In this section, we obtain an asymptotically optimal joint policy and analyze its performance. The main results are given as follows:
\begin{theorem}
	If the pdf $f(x)$ of the source symbols satisfies Conditions A and B, then the zero-wait sampling, the uniform quantization, and the real-valued AoI-optimal coding policies form an asymptotically optimal solution to the problem (\ref{joint optimal}), i.e., as the average distortion $D\to 0$, the combination achieves the optimal AoI asymptotically:
	\begin{align}
		\lim_{D\to 0} {\rm AoI}(S_{\mathrm{z}},Q_{\mathrm{uni}},F^{*})-{\rm AoI}(S^*,Q^*,F^*)=0.
	\end{align}
\end{theorem}
\begin{remark}
	In fact, we will see later that the proof of Theorem 1 implies that the zero-wait sampling, the uniform quantization, and the real-valued Shannon coding policies also form an asymptotically optimal solution.
\end{remark}

\begin{theorem}
	Under the jointly optimal policy, we have
	\begin{align} \label{slopethm}
		\lim_{\delta \to 0}\frac{{\rm AoI}(S_{\mathrm{z}},Q_{\mathrm{uni}},F^{*})}{\log_2 D}=-\frac{3}{4},
	\end{align}
	where $D$ denotes the average distortion and $\delta$ denotes the quantization cell size of the uniform quantization policy.
	Furthermore, we have
	\begin{align} \label{final}
		\lim_{\delta\to0}{\rm AoI}(S_{\mathrm{z}},Q_{\mathrm{uni}},F^{*})+\frac{3}{2}\log_2{\delta}=\frac{3}{2}h(X).
	\end{align}
\end{theorem}
\begin{remark}
	From equation (\ref{uniform3}),	
	we know that $\delta \to0$, asymptotically as the average distortion $D\to 0$.
\end{remark}
\begin{remark}
	In classical quantization theory, the performance curve of the entropy versus log distortion is a linear function with a slope of $-\frac{1}{2}$ asymptotically \cite[p.80]{gallager2008principles}. 
	Similarly,  the performance curve of the optimal AoI versus log distortion is a linear function with a slope of $-\frac{3}{4}$ asymptotically.  
\end{remark}
\begin{remark}
	In practical scenarios, the codeword lengths are integer-valued. Hence, we can choose the codeword lengths to be $\lceil l_i \rceil$, where $l_i$ are the obtained real-valued codeword lengths. For the uniform quantizer, we know that $H[Q_{{\mathrm{uni}}}(X)]>1$ for all sufficiently small $\delta$. Denoting $L':=\lceil L\rceil$, then we have
	\begin{align}
		&\frac{E[L'^2]}{2E[L']}+E[L']  \\\notag
		\le &\frac{E[(L+1)^2]}{2E[L]}+E[L+1]  \\\notag
		= &\frac{E[L^2]}{2E[L]}+\frac{1}{2E[L]}+E[L]+2 \\\notag
		\le &\frac{E[L^2]}{2E[L]}+E[L]+\frac{1}{2H[Q_{{\mathrm{uni}}}(X)]}+2 \\\notag
		< &\frac{E[L^2]}{2E[L]}+E[L]+\frac{5}{2},
	\end{align}
	which shows that there is a gap no greater than $\frac{5}{2}$ incurred by taking the integer constraint into account.
\end{remark}

We prove Theorem 1 in three steps:
we first show that the uniform quantizer is asymptotically optimal when the zero-wait sampler is given. We avoid solving the optimization problem (\ref{coding}) to get the optimal codeword lengths. Instead, we use the real-valued Shannon coding policy to approximate the solution of (\ref{coding}). 	
Then we prove that the zero-wait sampler is also asymptotically optimal when the uniform quantizer is given. 
Consequently, this combination achieves an asymptotically local minimum in the policy space.
Finally, we prove that this local minimum is in fact an asymptotically global minimum in the policy space.
In the end, we will provide the proof of Theorem 2. The key steps are sketched in the next three subsections.



\subsection*{A. Asymptotically Optimal Compressor for Zero-wait Sampler}
We first study the case where the zero-wait sampler is fixed. 
Given the zero-wait policy, we reformulate the problem (\ref{joint optimal}) as follows:
\begin{align}\label{quantization1}
	{\rm AoI}(S_{\mathrm{z}},Q^*,F^*)
	=&\min_{\{Q\in\mathcal{Q}\}}\min_{\{L\in\mathcal{L}\}}{\frac{E[L^2]}{2E[L]}+E[L]} \\ \notag
	&\mbox{s.t.} \sum_i 2^{-l_i}\le 1 \\\notag
	&\sum_i \int_{a_{i-1}}^{a_i} (x-c_i)^2 f(x)dx \le D \\\notag
	&\ l_i\in R^+.
\end{align}

It is worth noting that this tradeoff is similar to the entropy-distortion tradeoff in classical quantization theory \cite{Berger_1972}. However, this tradeoff is more complicated, since both calculating the optimal AoI and obtaining the real-valued AoI-optimal codeword lengths
require solving an optimization problem.
Thus it is difficult to obtain an exact solution to the problem (\ref{quantization1}).
However, we can get an asymptotically optimal solution in the high resolution regime. 
The result is stated as follows:
\begin{theorem}
	The uniform quantizer is asymptotically optimal when the zero-wait sampler is fixed.	
\end{theorem}
\begin{IEEEproof}
	We first present upper and lower bounds for ${\rm AoI}(S_{\mathrm{z}},Q^*,F^*)$ as follows:
	\begin{lemma}
		When the zero-wait sampler is given, the optimal AoI is bounded as 
		\begin{align}
			\frac{3}{2}H[Q^*(X)]\leq {\rm AoI}(S_{\mathrm{z}},Q^*,F^*) \leq {\rm AoI}(S_{\mathrm{z}},Q_{\mathrm{{\mathrm{uni}}}},F_\mathrm{s}).
		\end{align}   
	\end{lemma}
	\begin{IEEEproof}
		See Appendix A.
	\end{IEEEproof}
	
	The following lemma further shows that the difference between ${\rm AoI}(S_{\mathrm{z}},Q_{\mathrm{{\mathrm{uni}}}},F_\mathrm{s})$ and $\frac{3}{2}{H[Q_{{\mathrm{uni}}}(X)]}$  can be arbitrarily small by choosing $\delta$ small enough.
	\begin{lemma}
		If the zero-wait sampler and the uniform quantizer are given, then we have
		\begin{align}\label{lemma 3}
			\lim_{\delta \to 0}{\rm AoI}(S_{\mathrm{z}},Q_{{\mathrm{uni}}},F_\mathrm{s})-\frac{3}{2}{H[Q_{{\mathrm{uni}}}(X)]}= 0.
		\end{align}
	\end{lemma}
	\begin{IEEEproof}
		See Appendix B.
	\end{IEEEproof}
	
	Given any quantizer $Q'\in\mathcal{Q}$, by using Lemma 1 we obtain that for any $\epsilon >0$, there exists some $\delta_0 >0$, such that $0<\delta<\delta_0$ implies $\big|H[Q_{{\mathrm{uni}}}(X)]-H[Q^*(X)]\big|<\frac{\epsilon}{3}$ and hence
	\begin{align}
		&{\rm AoI}(S_{\mathrm{z}},Q',F^*)+\frac{\epsilon}{2}\\\notag
		\ge &\frac{3}{2}H[Q'(X)]+\frac{\epsilon}{2}  \\\notag
		\ge&\frac{3}{2}H[Q^*(X)]+\frac{\epsilon}{2}  \\\notag
		>&\frac{3}{2}H[Q_{{\mathrm{uni}}}(X)]  
	\end{align}
	where the first inequality follows from inequality (\ref{equ}). 
	
	By using Lemma 3, we obtain that for $\epsilon >0$, there exists some $\delta_1 >0$, such that $0<\delta<\delta_1$ implies $\big|{\rm AoI}(S_{\mathrm{z}},Q_{{\mathrm{uni}}},F_\mathrm{s})-\frac{3}{2}H[Q_{{\mathrm{uni}}}(X)]\big|<\frac{\epsilon}{2}$ and hence  
	\begin{align}
		{\rm AoI}(S_{\mathrm{z}},Q_{{\mathrm{uni}}},F^*)
		\le{\rm AoI}(S_{\mathrm{z}},Q_{{\mathrm{uni}}},F_\mathrm{s}) 
		<\frac{3}{2}H[Q_{{\mathrm{uni}}}(X)]+\frac{\epsilon}{2}.
	\end{align}
	Let $\delta'=\min\{\delta_0,\delta_1\}$, then for any $\delta$ satisfying  
	$0<\delta<\delta'$, then we have
	\begin{align}
		&{\rm AoI}(S_{\mathrm{z}},Q',F^*)+\epsilon   \\\notag
		>&\frac{3}{2}H[Q_{{\mathrm{uni}}}(X)]+\frac{\epsilon}{2}   \\\notag
		>&{\rm AoI}(S_{\mathrm{z}},Q_{{\mathrm{uni}}},F^*).	
	\end{align}
	Since $\epsilon$ can be arbitrarily small, we can obtain that 
	\begin{align}
		{\rm AoI}(S_{\mathrm{z}},Q',F^*)\ge{\rm AoI}(S_{\mathrm{z}},Q_{{\mathrm{uni}}},F^*).
	\end{align}
	Hence, the uniform quantizer is asymptotically optimal when the zero-wait sampler is given. This completes the proof.
\end{IEEEproof}
\subsection*{B. Asymptotically Optimal Sampler for Uniform Quantizer}
Here we will prove that the zero-wait sampler is asymptotically optimal when the uniform quantizer is given. We present the result as follows:
\begin{theorem}
	The zero-wait sampler is asymptotically optimal when the uniform quantizer is given.
\end{theorem}
\begin{IEEEproof}
	Suppose that the $S^*$ is the optimal sampling policy when the uniform quantizer is given, then we have
	\begin{align}
		&\quad \ {\rm AoI}(S^*,Q_{{\mathrm{uni}}},F^*)\\\notag 
		&=\frac{E[(L+Z(S^*))^2]}{2E[L+Z(S^*)]}+E[L]\\\notag
		&\ge \frac{1}{2}E[L+Z(S^*)]+E[L]\\\notag
		&\ge \frac{3}{2}E[L]\\\notag
		&\ge \frac{3}{2}H[Q_{{\mathrm{uni}}}(X)],
	\end{align}
	where the second-last inequality follows from the fact that $Z(S^*)\ge 0$.
	Then we can obtain upper and lower bounds directly for ${\rm AoI}(S^*,Q_{{\mathrm{uni}}},F^*)$ as follows:
	\begin{align}\label{bound2}
		&\frac{3}{2}H[Q_{{\mathrm{uni}}}(X)]\\\notag
		\le &{\rm AoI}(S^*,Q_{{\mathrm{uni}}},F^*) \\\notag
		\le &{\rm AoI}(S_{\mathrm{z}},Q_{{\mathrm{uni}}},F^*) \\\notag
		\le &{\rm AoI}(S_{\mathrm{z}},Q_{{\mathrm{uni}}},F_\mathrm{s}).
	\end{align} 
	We know that these bounds are asymptotically tight from Lemma 3. Hence, we only need to prove that the zero-wait sampling policy is asymptotically optimal when the uniform quantization and the real-valued Shannon coding policy are fixed. 
	
	According to  \cite[Corollary 1]{Sun_2019}, the zero-wait sampler is optimal if and only if for any $j$,
	\begin{align} \label{zero-wait condition1}
		E[{\rm ess}\inf L_j+L_{j+1}]\ge \frac{E[\int_{L_j}^{L_j+L_{j+1}}t dt]}{E[L_{j+1}]},
	\end{align}
	where $L_j$ represents the codeword length assigned to the $j$th symbol and ${\rm ess} \inf L_j=\inf\{l\in [0,\infty):P[L_j\leq l]>0\}$. 
	
	Thus we need to prove that the real-valued Shannon coding policy satisfies (\ref{zero-wait condition1}). By calculating the left and right sides of (\ref{zero-wait condition1}), we only need to verify that the real-valued Shannon coding policy satisfies the condition 
	\begin{align}\label{condition}
		{\rm ess}\inf L \ge \frac{E[L^2]}{2E[L]}.
	\end{align}
	For the real-valued Shannon coding policy, when the quantization cell size $\delta$ is small enough, we have ${\rm ess}\inf L\approx -\log_2{\delta M}$, where $M$ is the maximum of $f(x)$, as introduced in Condition A.
	From Lemma 1 and Lemma 3, we have  $\frac{1}{2}H[Q_{{\mathrm{uni}}}(X)]-\frac{1}{2}(h(X)-\log_2 \delta) \to 0$ and
	$\frac{E[L^2]}{2E[L]}-\frac{1}{2}H[Q_{{\mathrm{uni}}}(X)] \to 0 $ 
	as $\delta \to 0$.
	Thus we have
	\begin{align} \label{AoI and entropy}
		\frac{E[L^2]}{2E[L]}-\frac{1}{2}(h(X)-\log_2 \delta) \to 0  
	\end{align}  
	as $\delta \to 0$.  
	Clearly, condition (\ref{condition}) is satisfied when $\delta$ is small enough. 
	The details are provided in Appendix C.
\end{IEEEproof}
\begin{proposition}
	If the symbols satisfying Condition A and B are quantized uniformly and assigned the real-valued Shannon code, then the following holds:
	\begin{align}
		\lim_{\delta\to 0}\frac{E[L^2]}{E[L]^2}=1.
	\end{align}
\end{proposition}
\begin{IEEEproof}
	See Appendix D.
\end{IEEEproof}
\begin{remark}
	This property means that the codeword lengths are ``almost the same'' with their standard deviation negligible relative to their mean.
	From \cite{Sun_2019}, we also know that when the zero-wait policy is optimal, the service time is less random. Hence, this theorem is consistent with our intuition. 
\end{remark} 

In fact, for the integer-valued Shannon code, we also have a similar result: 
\begin{corollary}
	The zero-wait sampler is asymptotically optimal when the uniform quantizer and the integer-valued Shannon code are used.
\end{corollary}
\begin{IEEEproof}
	See Appendix E.
\end{IEEEproof}

So far, we obtain two locally optimal policies in an asymptotic sense from Theorem 3 and Theorem 4, i.e., $(S_\mathrm{z},Q_{{\mathrm{uni}}},F^*)$ and $(S_\mathrm{z},Q_{{\mathrm{uni}}},F_\mathrm{s})$. As we will see below, these policies are also globally asymptotically optimal.
\subsection*{C. Asymptotically Optimal Joint Sampling and Compression Policy}
Now, we present the proof of Theorem 1 as follows:
\begin{IEEEproof}
	Suppose that $(S^*,Q^*,F^*)$ is a globally optimal solution. By using the $(\epsilon,\delta)$-definition of the limit, we obtain that for any $\epsilon >0$, there exists some $\delta_0 >0$, such that $0<\delta<\delta_0$ implies $|H[Q_{{\mathrm{uni}}}(X)]-H[Q^*(X)]|<\frac{\epsilon}{3}$ and hence
	\begin{align}
		&\quad \ {\rm AoI}(S^*,Q^*,F^*)+\frac{\epsilon}{2}\\\notag 
		&=\frac{E[(L+Z(S^*))^2]}{2E[L+Z(S^*)]}+E[L]+\frac{\epsilon}{2}\\\notag
		&\ge \frac{1}{2}E[L+Z(S^*)]+E[L]+\frac{\epsilon}{2}\\\notag
		&\ge \frac{3}{2}E[L]+\frac{\epsilon}{2}\\\notag
		&\ge \frac{3}{2}H[Q^*(X)]+\frac{\epsilon}{2}\\\notag
		&> \frac{3}{2}H[Q_{{\mathrm{uni}}}(X)].
	\end{align}

	By using Lemma 3, we obtain that for any $\epsilon >0$, there exists some $\delta_1 >0$, such that $0<\delta<\delta_1$ implies $|{\rm AoI}(S_{\mathrm{z}},Q_{{\mathrm{uni}}},F_\mathrm{s})-\frac{3}{2}H[Q_{{\mathrm{uni}}}(X)]|<\frac{\epsilon}{2}$ and hence  
	\begin{align}
		{\rm AoI}(S_{\mathrm{z}},Q_{{\mathrm{uni}}},F^*)
		\le{\rm AoI}(S_{\mathrm{z}},Q_{{\mathrm{uni}}},F_\mathrm{s}) 
		<\frac{3}{2}H[Q_{{\mathrm{uni}}}(X)]+\frac{\epsilon}{2}.
	\end{align}
	Let $\delta'=\min\{\delta_0,\delta_1\}$, then for any $\delta$ satisfying  
	$0<\delta<\delta'$, then we have
	\begin{align}
		&{\rm AoI}(S^*,Q^*,F^*)+\epsilon \\\notag
		>& \frac{3}{2}H[Q_{{\mathrm{uni}}}(X)]+\frac{\epsilon}{2} \\\notag
		>& {\rm AoI}(S_\mathrm{z},Q_{{\mathrm{uni}}},F_\mathrm{s})\\\notag
		\ge& {\rm AoI}(S_{\mathrm{z}},Q_{{\mathrm{uni}}},F^*). 
	\end{align} 
	Since $\epsilon$ can be arbitrarily small, we conclude that
	\begin{align}
		{\rm AoI}(S^*,Q^*,F^*)\ge {\rm AoI}(S_{\mathrm{z}},Q_{{\mathrm{uni}}},F^*).
	\end{align} 
	Hence, the zero-wait sampler and the uniform quantizer are jointly optimal asymptotically. 
\end{IEEEproof}
Then we present the proof sketch of Theorem 2 as follows: 
\begin{IEEEproof}[Proof sketch of Theorem 2]
	We first prove (\ref{slopethm}). The average distortion $D$ will satisfy $\log_2 D < 0$ for all sufficiently small $\delta$.
	Then we have
	\begin{equation}
		\frac{\frac{3}{2}H[Q_{{\mathrm{uni}}}(X)]}{\log_2 D}\ge  \frac{{\rm AoI}(S_{\mathrm{z}},Q_{{\mathrm{uni}}},F^*)}{\log_2 D}\ge  \frac{{\rm AoI}(S_{\mathrm{z}},Q_{{\mathrm{uni}}},F_\mathrm{s})}{\log_2 D}.
	\end{equation}
	We hence obtain (\ref{slopethm}) by using a sandwich argument. The details are provided in Appendix F.
	
	Then we prove (\ref{final}). Similar to (\ref{AoI and entropy}), we have
	\begin{align} 
		{\rm AoI}(S_{\mathrm{z}},Q_{{\mathrm{uni}}},F^*)-\frac{3}{2}(h(X)-\log_2 \delta) \to 0
	\end{align}
	as $\delta \to 0$.
	The result hence follows immediately.
	The details are provided in Appendix F.
\end{IEEEproof}

\section{Numerical results}
\label{sec:submission}
In this section, we present numerical results to evaluate the performance of our proposed policies.
We plot AoI versus log distortion for different policies.
When we go along the curves from right to left, the number $N$ of quantization levels varies from 2 to 32. 

\begin{figure}[htbp]
	\centering
	\subfloat[$f(x)\sim \exp(1)$.]{%
		\label{fig:subfig1a}
		\includegraphics[width=8cm]{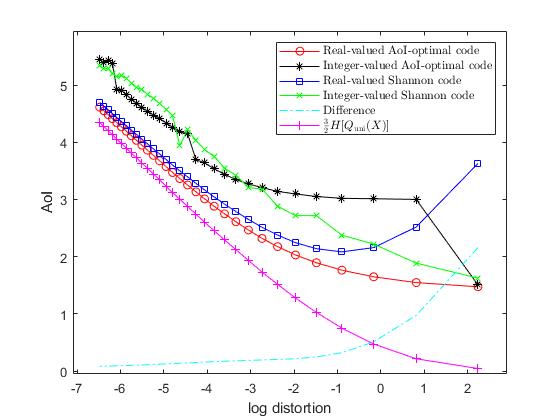}}
	\hfil
	\subfloat[$f(x)\sim N(0,1)$.]{%
		\label{fig:subfig1b}
		\includegraphics[width=8cm]{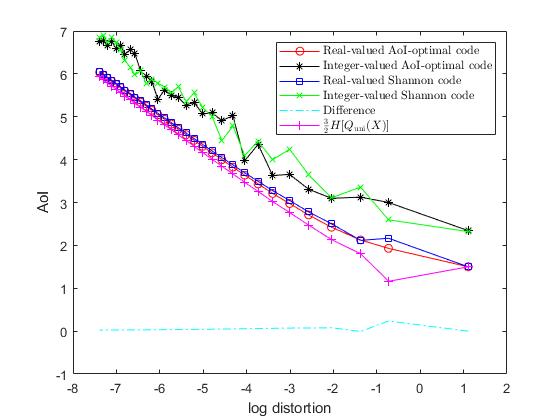}}  
	\caption{Performance curves of the AoI versus log distortion with different coding policies when the zero-wait sampling and uniform quantization policies are fixed for symbols with pdfs $f(x)\sim \exp(1)$ and $f(x)\sim N(0,1)$.}
	\label{fig:subfigures1}
\end{figure}

\figurename \ \ref{fig:subfigures1} illustrates the performance of the real-valued and integer-valued Shannon coding policies as well as the real-valued and  integer-valued AoI-optimal coding policies for source symbols with pdfs $f(x)\sim\exp(1)$ and $f(x)\sim N(0,1)$ when the zero-wait sampler and the uniform quantizer are given. We truncate the pdfs $f(x)\sim\exp(1)$ and $f(x)\sim N(0,1)$ to $[0, 15]$ and $[-5, 5]$ respectively, so as to meet Conditions A and B.   Moreover, we plot the difference of the corresponding AoI for the two real-valued coding policies, and also plot $\frac{3}{2}H[Q_{{\mathrm{uni}}}(X)]$ as a lower bound.

As we can observe, there is a small gap between the AoI of the real-valued AoI-optimal code and that of the real-valued Shannon code when the zero-wait sampler and the uniform quantizer are fixed. The gap is asymptotically close to zero as the number of quantization levels grows. At the same time, we find that both the curves are asymptotically linear.
Moreover, we observe that the three curves corresponding to $\frac{3}{2}H[Q_{{\mathrm{uni}}}(X)]$ and the AoI of the two real-valued
coding policies
have a strong tendency to overlap, thereby confirming Lemma 3. Furthermore, the slope of these curves are equal to $-\frac{3}{4}$, confirming Theorem 2. We also observe that there is a gap no greater than $\frac{5}{2}$ incurred by using ceiling function for the two real-valued coding policies and that the gap between the two integer-valued coding policies is small.

\begin{figure}[htbp]
	\centering
	\subfloat[$f(x)\sim\exp(1)$.]{%
		\label{fig:subfig2a}
        \includegraphics[width=8cm]{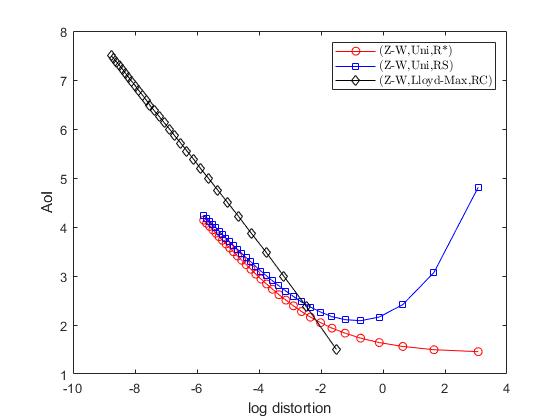}}
	\hfil
	\subfloat[$f(x)\sim N(0,1)$.]{%
		\label{fig:subfig2b}
		\includegraphics[width=8cm]{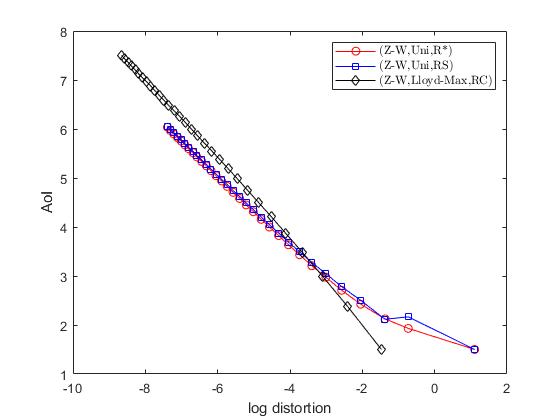}}  
	\caption{Performance curves of the AoI versus log distortion with different quantization and real-valued coding policies for symbols with pdfs $f(x)\sim\exp(1)$ and $f(x)\sim N(0,1)$.}
	\label{fig:subfigures2}
\end{figure}

\begin{figure}[htbp]
	\subfloat[$f(x)\sim\exp(1)$.]{%
		\label{fig:subfig3a}
         \includegraphics[width=8cm]{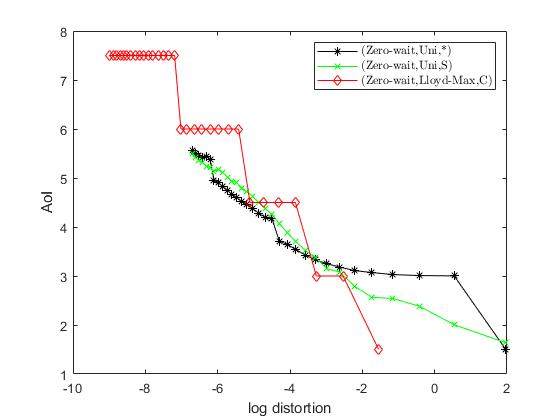}}
	\hfil
	\subfloat[$f(x)\sim N(0,1)$.]{%
		\label{fig:subfig3b}
		\includegraphics[width=8cm]{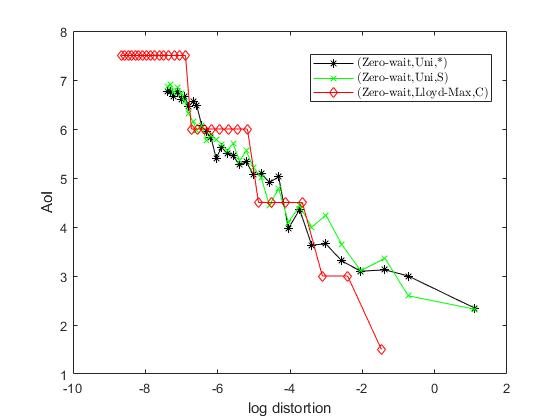}}  
	\caption{Performance curves of the AoI versus log distortion with different quantization and integer-valued coding policies for symbols with pdfs $f(x)\sim\exp(1)$ and $f(x)\sim N(0,1)$.}
	\label{fig:subfigures4}
\end{figure}

In \figurename \ \ref{fig:subfigures2} and \figurename \ \ref{fig:subfigures4}, we compare the performance between the uniform quantizer and the Lloyd-Max quantizer followed by a constant length code when the zero-wait sampler is given.    
We use ``Uni'' and ``Lloyd-Max'' to represent the uniform quantizer and the Lloyd-Max quantizer respectively. Moreover, the legends ``R*'', ``RS'' and ``RC'' represent the real-valued AoI-optimal code, the real-valued Shannon code, and the code for the Lloyd-Max quantizer with a constant codeword length $\log_2 N$. In addition, the legends ``*'', ``S'' and ``C'' represent the integer-valued AoI-optimal code, the integer-valued Shannon code, and the code for the Lloyd-Max quantizer with a constant codeword length $\lceil \log_2 N\rceil$.

In \figurename \ \ref{fig:subfig2a}, we present the results for pdf $f(x)\sim\exp(1)$. We can see that the Lloyd-Max quantizer provides a lower distortion compared to the uniform quantizer, but the corresponding AoI grows very quickly. In \figurename \ \ref{fig:subfig2b}, we present the results for pdf $f(x)\sim N(0,1)$. Although the gap between two quantizers is not as large as that in \figurename \ \ref{fig:subfig2a}, we still observe that the uniform quantizer provides a lower AoI and the gap is getting wider as the resolution becomes finer. This corroborates Theorem 1.
In \figurename \ \ref{fig:subfigures4}, when we take the integer constraint into account, we observe that there is a larger gap between two quantizers for the great majority of $N$.
These results imply that the Lloyd-Max quantizer does not always perform well in real-time applications.

%



\section{Conclusion}
In this work, we consider a problem of designing joint sampling, quantization and coding policy to minimize the AoI subject to the distortion constraint of the samples in a status update system with single generate-at-will source. We prove that the zero-wait sampler, uniform quantizer, and the real-valued AoI-optimal code are asymptotically optimal. Moreover, the performance curve of the optimal AoI versus log distortion is a linear function with a slope of $-\frac{3}{4}$ asymptotically. We also show that the real-valued Shannon code can achieve the optimal performance asymptotically.

\enlargethispage{-1.4cm}


\bibliography{bibliography}

\appendices

\section{Proof of Lemma 2}
When a quantizer $Q$ is given, by using Jensen's inequality $E[L^2]\geq E[L]^2$ and the fact that $E[L]\geq H[Q(X)]$ for a prefix-free code, then we have
\begin{align}
	\frac{E[L^2]}{2E[L]}+E[L]\geq \frac{3}{2}E[L]\geq \frac{3}{2}H[Q(X)].
\end{align}
Hence we have
\begin{align}\label{equ}
	{\rm AoI}(S_{\mathrm{z}},Q,F^*) \ge \frac{3}{2}H[Q(X)].
\end{align}
Then we have
\begin{align}
	{\rm AoI}(S_{\mathrm{z}},Q^*,F^*) \ge \frac{3}{2}H[Q^*(X)].
\end{align}     
The lower bound is proved. Then we can directly obtain the upper bound by using the definition of the optimality.

\section{Proof of Lemma 3}
%
%

If the the quantization cell size $\delta$ is small enough, then the $f(x)$ will be approximately constant over each cell, and the occurrence probability of the $i$th representation point is
\begin{align} \label{pro}
	p_i\approx f(x_i)\delta.
\end{align}
We assign the real-valued Shannon code to the quantized symbol, i.e., $l_i=-\log_2 p_i$, we have
\begin{align} \label{secondE}
	E[L^2]=\sum_i p_i\log_2^2 {p_i}.
\end{align}
By substituting (\ref{pro}) into (\ref{secondE}), we have
\begin{align}
	E[L^2]=&\sum_i f(x_i)\delta \big(\log_2 {\big(f(x_i)\delta\big)}\big)^2   \\ \notag
	=&\sum_i f(x_i)\delta \big(\log_2 {f(x_i)}+\log_2 {\delta}\big)^2   \\ \notag
	=&\sum_i f(x_i)\delta\big(\log_2 ^2{f(x_i)}+2\log_2 {f(x_i)}\log_2{\delta}\big)+\log_2 ^2{\delta}.
\end{align}
Similarly, we can obtain
\begin{align}
	E[L]&=-\sum_i f(x_i)\delta \log_2 {\big(f(x_i)\delta\big)}\\ \notag
	&=-\sum_i f(x_i)\delta \log_2 {f(x_i)}-\log_2{\delta}.
\end{align}
When $\delta \to 0$, we have
\begin{align} \label{h}
	-\sum_i f(x_i)\delta \log_2 {\big(f(x_i)\big)}\longrightarrow h(X)
\end{align}
and
\begin{align} \label{var}
	\sum_i f(x_i)\delta\big(\log_2 ^2{f(x_i)}\big)\longrightarrow \int_{I}f(x)\log_2^2 {\big(f(x)\big)}dx.
\end{align}
Thus we have
\begin{align}
	&{\rm AoI}(S_{\mathrm{z}},Q_{{\mathrm{uni}}},F_\mathrm{s})-\frac{3}{2}{H[Q_{{\mathrm{uni}}}(X)]}\\\notag
	\\\notag
	=&\frac{E[L^2]}{2E[L]}+E[L]-\frac{3}{2}{H[Q_{{\mathrm{uni}}}(X)]}\\\notag
	\\\notag
	=&-\frac{\sum_i f(x_i)\delta \log_2^2 {\bigg(f(x_i)\delta\bigg)}}{2\sum_i f(x_i)\delta \log_2 {\bigg(f(x_i)\delta\bigg)}}+\frac{1}{2}\sum_i f(x_i)\delta \log_2 {\bigg(f(x_i)\delta\bigg)}\\ \notag
	\\\notag
	=&\frac{-\bigg(\sum_i f(x_i)\delta\bigg(\log_2 ^2{f(x_i)}+2\log_2 {f(x_i)}\log_2{\delta}\bigg)+\log_2 ^2{\delta}\bigg)}{2\bigg(\sum_i f(x_i)\delta \log_2 {f(x_i)}+\log_2{\delta}\bigg)}\\\notag
	+&\frac{\bigg(\sum_i f(x_i)\delta \log_2 {f(x_i)}+\log_2{\delta}\bigg)^2}{2\bigg(\sum_i f(x_i)\delta \log_2 {f(x_i)}+\log_2{\delta}\bigg)}\\\notag
	\\\notag
	=&\frac{-\sum_i f(x_i)\delta \log_2 ^2{f(x_i)}+\bigg(\sum_i f(x_i)\delta \log_2 {f(x_i)}\bigg)^2}{2\bigg(\sum_i f(x_i)\delta \log_2 {f(x_i)}+\log_2{\delta}\bigg)}.
\end{align}
Thus we have
\begin{equation}
	\lim_{\delta \to 0}{\rm AoI}(S_{\mathrm{z}},Q_{{\mathrm{uni}}},F_\mathrm{s})-\frac{3}{2}{H[Q_{{\mathrm{uni}}}(X)]}= 0.
\end{equation}

\section{Proof of Theorem 4}

Suppose that the $S^*$ is the optimal sampling policy when the uniform quantizer is given, then we have
\begin{align}
	&\quad \ {\rm AoI}(S^*,Q_{{\mathrm{uni}}},F^*)\\\notag 
	&=\frac{E[(L+Z(S^*))^2]}{2E[L+Z(S^*)]}+E[L]\\\notag
	&\ge \frac{1}{2}E[L+Z(S^*)]+E[L]\\\notag
	&\ge \frac{3}{2}E[L]\\\notag
	&\ge \frac{3}{2}H[Q_{{\mathrm{uni}}}(X)],
\end{align}
where the second-last inequality follows from the fact that $Z(S^*)\ge 0$.
Then we can obtain upper and lower bounds directly for ${\rm AoI}(S^*,Q_{{\mathrm{uni}}},F^*)$ as follows:
\begin{align}\label{bound2}
	&\frac{3}{2}H[Q_{{\mathrm{uni}}}(X)]\\\notag
	\le &{\rm AoI}(S^*,Q_{{\mathrm{uni}}},F^*) \\\notag
	\le &{\rm AoI}(S_{\mathrm{z}},Q_{{\mathrm{uni}}},F^*) \\\notag
	\le &{\rm AoI}(S_{\mathrm{z}},Q_{{\mathrm{uni}}},F_\mathrm{s}).
\end{align} 
We know that these bounds are asymptotically tight from Lemma 3. Hence, we only need to prove that the zero-wait sampling policy is asymptotically optimal when the uniform quantization and the real-valued Shannon coding policy are fixed. 

According to  \cite[Corollary 1]{Sun_2019}, the zero-wait sampler is optimal if and only if for any $j$,
\begin{align} \label{zero-wait condition}
	E[{\rm ess}\inf L_j+L_{j+1}]\ge \frac{E[\int_{L_j}^{L_j+L_{j+1}}t dt]}{E[L_{j+1}]},
\end{align}
where $L_j$ represents the codeword length assigned to the $j$th symbol and ${\rm ess} \inf L_j=\inf\{l\in [0,\infty):P[L_j\leq l]>0\}$. Because the source symbols are i.i.d., we have $E[L_j]=E[L_{j+1}]=E[L]$. 
The right side of (\ref{zero-wait condition}) is
\begin{align}
	\frac{E[\int_{L_j}^{L_j+L_{j+1}}t dt]}{E[L_{j+1}]}
	=\frac{E[L^2]}{2E[L]}+E[L].  
\end{align}
The left side of (\ref{zero-wait condition}) is
\begin{align}
	E[{\rm ess}\inf L_j+L_{j+1}]={\rm ess}\inf L_j+E[L].
\end{align}
If $\delta$ is small enough, then $f(x)$ will be approximately constant over each cell, and we have 
\begin{align}
	{\rm ess}\inf L_j\approx -\log_2{\delta M}.
\end{align}
where $M$ is the maximum of the $f(x)$.
By using Lemma 1 and the $(\epsilon,\delta)$-definition of limit, then we have
for any $\epsilon >0$, there exists some $\delta_0 >0$,
such that for any $\delta$ satisfying  
$0<\delta<\delta_0$, it holds that 
\begin{align}
	\bigg|\frac{1}{2}{H[Q_{{\mathrm{uni}}}(X)]}+\frac{1}{2}\log_2{\delta}-\frac{1}{2}h(X)\bigg|<\frac{\epsilon}{2}.
\end{align}
Similarly, when the symbols are quantized uniformly and assigned the real-valued Shannon code, there exists some $\delta_1 >0$, such that for any $\delta$ satisfying  $0<\delta<\delta_1$, it holds that
\begin{align}
	\bigg|\frac{E[L^2]}{2E[L]}-\frac{1}{2}{H[Q_{{\mathrm{uni}}}(X) ]}\bigg|<\frac{\epsilon}{2}.
\end{align}
Let $\delta'=\min\{\delta_0,\delta_1\}$, then for any $\delta$ satisfying  
$0<\delta<\delta'$, we have
\begin{align}
	&\bigg|\frac{E[L^2]}{2E[L]}+\frac{1}{2}\log_2{\delta}-\frac{1}{2}h(X)\bigg|\\\notag
	<&\bigg|\frac{E[L^2]}{2E[L]}-\frac{1}{2}{H[Q_{{\mathrm{uni}}}(X)]}\bigg| \\\notag
	+&\bigg|\frac{1}{2}{H[Q_{{\mathrm{uni}}}(X)]}+\frac{1}{2}\log_2{\delta}-\frac{1}{2}h(X)\bigg|\\\notag
	<&\frac{\epsilon}{2}+\frac{\epsilon}{2}\\\notag
	=&\epsilon. 
\end{align}
Let $\delta_2=2^{-2\log_2M-h(X)-2\epsilon-1}$ and $\delta''=\min\{\delta_2,\delta '\}$, then for any $\delta$ satisfying  
$0<\delta<\delta''$, we have
\begin{align}
	&E[{\rm ess}\inf L_j+L_{j+1}]-\frac{E[\int_{L_j}^{L_j+L_{j+1}}t dt]}{E[L_{j+1}]}\\\notag
	=&-\log_2{\delta M}-\frac{E[L^2]}{2E[L]}	\\\notag
	>&-\log_2{\delta M}-\frac{1}{2}h(X)+\frac{1}{2}\log_2{\delta}-\epsilon \\\notag
	=&-\frac{1}{2}\log_2{\delta}-\log_2{M}-\frac{1}{2}h(X)-\epsilon \\\notag
	>&0.
\end{align}
This completes the proof.

\section{Proof of Proposition 1}
Like the proof of Lemma 3, we have
\begin{align}
	&\lim_{\delta\to 0}\frac{E[L^2]}{E[L]^2}\\\notag
	=&\lim_{\delta\to 0}\frac{\sum_i f(x_i)\delta \log_2^2 {(f(x_i)\delta)}}{\bigg(\sum_i f(x_i)\delta \log_2 {\big(f(x_i)\delta\big)}\bigg)^2}\\\notag
	=&\lim_{\delta\to 0}\frac{\sum_i f(x_i)\delta\bigg(\log_2 ^2{f(x_i)}+2\log_2 {f(x_i)}\log_2{\delta}\bigg)+\log_2 ^2{\delta}}{\bigg(\sum_i f(x_i)\delta \log_2 {f(x_i)}+\log_2{\delta}\bigg)^2}\\\notag
	=&1.
\end{align}
\section{Proof of Corollary 1}
For the uniform quantizer, we know that $H[Q_{{\mathrm{uni}}}(X)]>1$ for all sufficiently small $\delta$. Then we have
\begin{align}
	&\frac{E[L'^2]}{2E[L']}  \\\notag
	\le &\frac{E[(L+1)^2]}{2E[L]}  \\\notag
	\le &\frac{E[L^2]}{2E[L]}+\frac{1}{2H[Q_{{\mathrm{uni}}}(X)]}+1 \\\notag
	< &\frac{E[L^2]}{2E[L]}+\frac{3}{2}.
\end{align}
We only need to verify that the integer-valued Shannon code satisfies the condition (\ref{condition}).
Then we have
\begin{align}
	&{\rm ess}\inf L' - \frac{E[L'^2]}{2E[L']}\\\notag
	=&\lceil-\log_2{\delta M}\rceil-\frac{E[L'^2]}{2E[L']}	\\\notag
	\ge&-\log_2{\delta M}-1-\frac{E[L^2]}{2E[L]}-\frac{3}{2} \\\notag
	\ge&-\log_2{\delta M}-1-\frac{1}{2}h(X)+\frac{1}{2}\log_2{\delta}-\frac{3}{2}-\epsilon \\\notag
	=&-\frac{1}{2}\log_2{\delta}-\log_2{M}-\frac{1}{2}h(X)-\frac{5}{2}-\epsilon \\\notag
	>&0.
\end{align}
Asymptotically for very small $\delta$, the last inequality holds. This completes the proof.

\section{Proof of Theorem 2}
The average distortion $D$ will satisfy $\log_2 D < 0$ for all sufficiently small $\delta$.
Then we have
\begin{equation}
	\frac{\frac{3}{2}H[Q_{{\mathrm{uni}}}(X)]}{\log_2 D}\ge  \frac{{\rm AoI}(S_{\mathrm{z}},Q_{{\mathrm{uni}}},F^*)}{\log_2 D}\ge  \frac{{\rm AoI}(S_\mathrm{z},Q_{{\mathrm{uni}}},F_\mathrm{s})}{\log_2 D}.
\end{equation}
By using Lemma 1, we have
\begin{equation}\label{slope}
	\lim_{\delta \to 0}\frac{\frac{3}{2}H[Q_{{\mathrm{uni}}}(X)]}{\log_2 D}=-\frac{3}{4}.
\end{equation}
Lemma 3 implies that 
\begin{align}\label{slopeAoI}
	\lim_{\delta \to 0}\frac{{\rm AoI}(S_{\mathrm{z}},Q_{{\mathrm{uni}}},F_\mathrm{s})}{H[Q_{{\mathrm{uni}}}(X)]}
	=\frac{3}{2}.
\end{align}
It follows that
\begin{align}
	&\lim_{\delta \to 0}\frac{{\rm AoI}(S_{\mathrm{z}},Q_{\mathrm{uni}},F_\mathrm{s})}{\log_2 D} \\\notag
	=&\lim_{\delta \to 0}\frac{{\rm AoI}(S_{\mathrm{z}},Q_{\mathrm{uni}},F_\mathrm{s})}{H[Q_{\mathrm{uni}}(X)]}\frac{H[Q_{\mathrm{uni}}(X)]}{\log_2 D} \\\notag
	=&-\frac{3}{4}.
\end{align}
By using a sandwich argument, we have
\begin{equation}
	\lim_{\delta \to 0}\frac{{\rm AoI}(S_{\mathrm{z}},Q_{\mathrm{uni}},F^*)}{\log_2 D}=-\frac{3}{4}.
\end{equation}
Furthermore, 
\begin{align}
	&\bigg|{\rm AoI}(S_{\mathrm{z}},Q_{\mathrm{uni}},F^*)+\frac{3}{2}\log_2{\delta}-\frac{3}{2}h(X)\bigg|\\\notag
	\leq&\bigg|{\rm AoI}(S_{\mathrm{z}},Q_{\mathrm{uni}},F^*)-\frac{3}{2}H[Q_{\mathrm{uni}}(X)]\bigg|\\\notag
	+&\bigg|\frac{3}{2}H[Q_{\mathrm{uni}}(X)]+\frac{3}{2}\log_2{\delta}-\frac{3}{2}h(X)\bigg|.\notag
\end{align}
We know that
\begin{align}
	\lim_{\delta \to 0} {\rm AoI}(S_{\mathrm{z}},Q_{{\mathrm{uni}}},F^*)-\frac{3}{2}H[Q_{\mathrm{uni}}(X)]=0
\end{align}
and
\begin{align}
	\lim_{\delta \to 0}\frac{3}{2}H[Q_{\mathrm{uni}}(X)]+\frac{3}{2}\log_2{\delta}-\frac{3}{2}h(X)=0.
\end{align}
Then we have
\begin{align}
	\lim_{\delta \to 0}{\rm AoI}(S_{\mathrm{z}},Q_{\mathrm{uni}},F^*)+\frac{3}{2}\log_2{\delta}=\frac{3}{2}h(X).     
\end{align}
This completes the proof.

\end{document}